\documentclass[pre,amssymb,amsmath]{revtex4}
\usepackage{graphics}
\usepackage{graphics}
\usepackage{epsfig}
\usepackage{color}
\def\dd{{\mathrm d}}
\def\Jav{\langle J \rangle}

\begin{document}

\title{Statistical mechanics and Vlasov equation allow for a simplified
Hamiltonian description of Single-Pass Free Electron Laser
saturated dynamics}

\author{Andrea Antoniazzi$^{1}$\thanks{antoniazzi@docs.de.unifi.it},Yves Elskens$^{2}$\thanks{Yves.Elskens@up.univ-mrs.fr}, \\
Duccio Fanelli$^{1,3}$\thanks{fanelli@et3.cmb.ki.se},
 Stefano Ruffo$^{1}$\thanks{stefano.ruffo@unifi.it}\\
}
\affiliation{1.Dipartimento di Energetica, Universit{\`a} di Firenze and
INFN, via S. Marta, 3, 50139 Firenze, Italy\\
2. Equipe Turbulence Plasma de l'UMR 6633 CNRS--Universit\'e de Provence,
case 321, campus Saint-J\'er{\^{o}}me, F-13397 Marseille cedex 13,
France \\
3. Department of Cell and Molecular Biology, Karolinska Institute,
SE-171 77 Stockholm, Sweden\\
}

\date{\today}

\begin{abstract}
A reduced Hamiltonian formulation to reproduce the saturated regime of a Single
 Pass Free Electron Laser, around perfect tuning, is here discussed. Asymptotically,
 $N_m$ particles are found to organize in a dense cluster, that evolves as
 an individual massive unit. The remaining particles fill the surrounding
 uniform sea, spanning a finite portion of phase space, approximately delimited
 by the average momenta $\omega_+$ and $\omega_-$. These quantities enter
 the model as external parameters, which can be self-consistently determined
 within the proposed theoretical framework. To this aim, we make use of a
 statistical mechanics treatment of the Vlasov equation, that governs the
 initial amplification process. Simulations of the reduced dynamics are shown
 to successfully capture the oscillating regime observed within the
 original $N$-body picture.

\end{abstract}

\maketitle

%*******************************************************************
\section{General background}
\label{sec1}
%*******************************************************************

Free-Electron Lasers (FELs) are coherent and tunable
radiation sources, which differ from conventional lasers in using a
relativistic electron beam as their lasing medium, hence the term free-electron.

The physical mechanism responsible for the light emission and
amplification is the interaction  between the relativistic electron
beam, a magnetostatic periodic field generated in the
undulator and an optical wave copropagating with the
electrons. Due to the effect of the magnetic  field, the electrons are forced to follow
sinusoidal trajectories, thus emitting synchrotron radiation. This
{\it spontaneous emission} is then amplified along the undulator until the laser effect is reached.
Among different schemes, single-pass high-gain FELs are currently attracting growing
interest, as they are promising sources of
powerful and coherent light in the UV and X
ranges. Besides the Self Amplified Spontaneous
Emission (SASE) setting \cite{sase},  seeding schemes may be adopted where a
small laser signal is injected at the entrance of the undulator
and guides the subsequent amplification process \cite{seed}. In the following we
shall refer to the latter case. Basic features of the system dynamics are successfully captured by a simple one-dimensional
Hamiltonian model \footnote{Note that the model here considered does not account
for the spectral properties of the radiation
nor include the effect of the slippage, i.e. the velocity difference
between the electron beam and the co-propagating wave. More detailed
formulations are to be invoked to achieve a full description
of the laser performance at the exit of the undulator.}
introduced by Bonifacio and collaborators in
\cite{Bonifacio}. Remarkably, this simplified formulation applies
to other physical systems, provided a formal translation of the
variables involved is performed. As an example, focus on kinetic plasma
turbulence, e.g. \ the electron beam-plasma instability.
When a weak electron beam is injected into a thermal plasma,
electrostatic modes at the plasma frequency (Langmuir modes) are
destabilized.  The interaction of the Langmuir waves
and the electrons constituting the beam can be studied in the framework
of a self-consistent Hamiltonian picture \cite{ElskensBook}, formally equivalent to
the one in \cite{Bonifacio}. In a recent paper \cite{andrea} we established
a bridge between these two areas of investigation (FEL and
plasma), and exploited the connection to derive a reduced Hamiltonian
model to characterize the saturated dynamics of the laser.
According to this scenario, $N_m$ particles are trapped in the
resonance, i.e. experience a bouncing motion
in one of the (periodically repeated) potential wells,
and form a clump that evolves as a single macro-particle localized in space.
The remaining particles populate the surrounding halo, being almost
uniformly distributed in phase space between two sharp boundaries,
whose average momentum is labeled $\omega_+$ and $\omega_-$.
The issue of providing a self-consistent estimate
for the external parameters $N_m$, $\omega_+$ and $\omega_-$ is
addressed and solved in this paper.

%Let us emphasize that
This long-standing problem was first pointed out by
Tennyson et al. in the pioneering work \cite{tennyson}
and recently revisited in \cite{ElskensBook}. A first attempt to calculate
$N_m$ is made in \cite{tesi} where a semi-analytical argument is proposed.
In this respect, the strategy here proposed applies to a large class
of phenomena whose dynamics can be modeled within a Hamiltonian
framework \cite{ElskensBook,Ruffo} displaying the emergence of
collective behaviour \cite{Kaneko}.

The paper is organized as follows. In Section \ref{sec2} we
introduce the one-dimensional model of a FEL amplifier
\cite{Bonifacio} and review the derivation of the reduced Hamiltonian
\cite{tennyson,andrea}. Section  \ref{sec3} recalls the statistical
mechanics approach to estimate the saturated laser regime. In Sections
\ref{sec4} to \ref{sec6} the analytic characterization
of $N_m$, $\omega_+$ and $\omega_-$ is given in details and the results
are then tested numerically in section \ref{sec7}. Finally, in Section \ref{sec8} we sum
up and draw our conclusions.

%*******************************************************************
\section{From the self-consistent $N$-body Hamiltonian to the reduced formulation}
\label{sec2}
%*******************************************************************

Under the hypothesis of one-dimensional motion
and monochromatic radiation, the steady state
dynamics of a Single-Pass Free Electron Laser is described by the
following set of equations:

\begin{eqnarray}
\frac{{\mathrm d}\theta_j}{{\mathrm d}\bar{z}} &=& p_j\quad, \label{eq:mvttheta}\\
\frac{{\mathrm d}p_j}{{\mathrm d}\bar{z}} &=&
-Ae^{i\theta_j}-A^{\ast}e^{-i\theta_j}\quad,
\label{eq:mvtp}\\
\frac{{\mathrm d}A}{{\mathrm d}\bar{z}} &=&
i \delta A +\frac{1}{N} \sum_j e^{-i\theta_j}~\quad, \label{eq:mvtA}
\end{eqnarray}
where $\bar{z}=2k_w \rho z \gamma_r^2/\langle \gamma \rangle_0^2$ is the
rescaled longitudinal coordinate, which plays the role of time.
Here, $\rho=[a_{w}\omega_p/(4ck_w)]^{2/3}/\gamma_r$ is the so-called
Pierce parameter, $\langle \gamma
\rangle_0$ the mean energy of the electrons at the undulator's
entrance, $k_w=2 \pi / \lambda_w$ the wave
number of the undulator, $\omega_p=(4 \pi e^2n/m)^{1/2}$ the plasma frequency, $c$
the speed of light, $n$ the total electron number density,
$e$ and  $m$ respectively the charge and mass
of one electron. Further, $a_w=eB_w/(k_w m c^2)$, where $B_w$ is the
rms peak undulator field.
Here $\gamma_r=\left( \lambda_w
(1+a_w^2)/2 \lambda \right)^{1/2}$ is the resonant energy,  $\lambda_w$ and $\lambda$
being respectively the period of the undulator and the wavelength of
the radiation field. Introducing  the wavenumber $k$ of the FEL
radiation, the two canonically conjugated variables are
($\theta$,$p$), defined as
$\theta=(k+k_w)z-2\delta\rho k_w z \gamma_r^2/\langle \gamma
\rangle_0^2$ and $p=(\gamma-\langle \gamma
\rangle_0)/(\rho \langle \gamma \rangle_0)$. $\theta$ corresponds to the phase
of the electrons with respect to the ponderomotive wave. The complex amplitude $A=A_x + iA_y $
represents the
scaled field, transversal to $z$.  Finally, the detuning
parameter is given by $\delta=(\langle \gamma
\rangle_0^2-\gamma_r^2)/(2\rho\gamma_r^2)$, and measures the
average relative deviation from the resonance condition.

The above system of equations ($N$ being the number of electrons) can be derived from the Hamiltonian
\begin{equation}
H=\sum_{j=1}^N\frac{p_j^2}{2} -\delta I + 2\sqrt{\frac{I}{N}}\sum_{j=1}^N
\sin(\theta_j-\varphi) \label{eq:Hamiltonien},
\end{equation}
where the intensity~$I$ and the
phase~$\varphi$ of the wave are given by
$A=\sqrt{I/N} \exp(-i \varphi)$. Here the canonically conjugated variables are
$(p_j,\theta_j)$ for $1 \leq j \leq N$ and $(I,\varphi)$.
Besides the ``energy'' $H$, the total momentum $P=\sum_j
p_j +  I$ is also conserved.
By exploiting these conserved quantities, one can recast the FEL
equations of motion in the following form for the set of
$2N$ conjugate variables ($q_j,p_j$) \cite{farina}:

\begin{eqnarray}
\label{eqs1}
\dot{q_j} &=&  p_j - \frac{1}{\sqrt{NI}}\sum_{l=1}^{N} \sin q_l  +\delta ~, \label{pfix0} \\
\dot{p_j} &=& - 2 \sqrt{\frac{I}{N}} \cos q_j \label{qfix0},
\end{eqnarray}
where the dot denotes derivation with respect to $\bar{z}$, and $q_j=\theta_j - \varphi \ {\mathrm{mod}} (2 \pi)$  is the
 phase of the $j^{th}$ electron in a proper reference
frame. The fixed points of system
(\ref{eqs1})-(\ref{qfix0}) are
determined by imposing
$\dot{q_j}=\dot{p_j}=0$ and solving:

\begin{eqnarray}
p_j - \frac{1}{\sqrt{NI}}\sum_{l=1}^{N} \sin q_l + \delta&=& 0 ~,   \label{pfix}\\
2\sqrt{\frac{I}{N}} \cos q_j &=& 0 \label{qfix}.
\end{eqnarray}
An elliptical fixed point is found for $q_i=\bar{q}=3\pi/2$.
The conjugate momentum solves $({\bar p}+\delta)
\sqrt{P/N - {\bar p}} + 1 = 0$ and therefore depends on $P/N$.
We shall return on this issue in the following Sections.

\begin{figure}[ht]
\vspace{1truecm}
 \resizebox{0.6\textwidth}{!}{\includegraphics{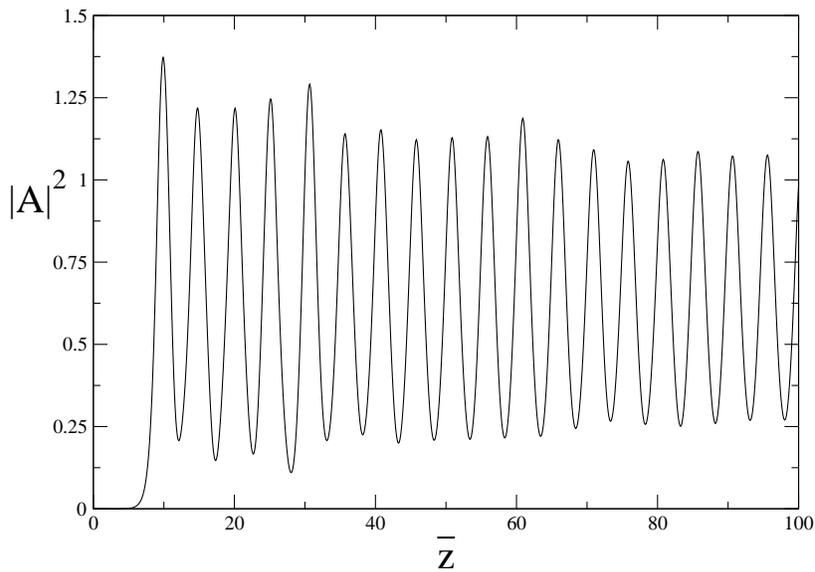}}
\vskip .5truecm \caption{\em Evolution of the radiation
intensity as follows from equations   (\ref{eq:mvttheta}),
(\ref{eq:mvtp}) and (\ref{eq:mvtA}).
$N=10^4$ electrons are simulated, for an initial
mono-energetic profile. Here $\delta=0$ and
$I(0) \simeq 0$.
Particles are initially uniformly
distributed in space.
\label{fig1}}
\end{figure}

\begin{figure}[hb]
\vspace{1truecm}
 \resizebox{0.64\textwidth}{!}{\includegraphics{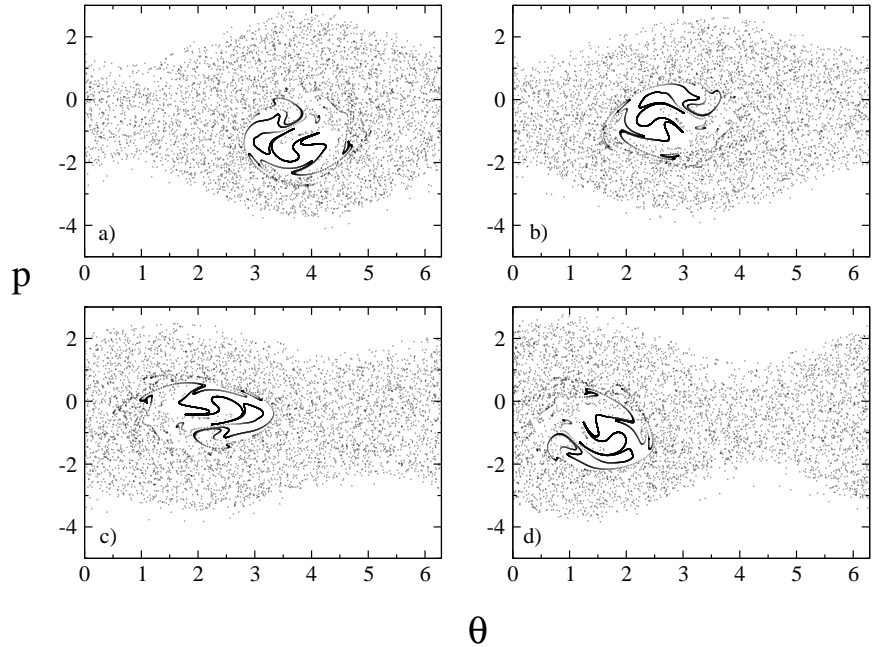}}
\vskip .5truecm \caption{\em
Phase space portraits for different position along the undulator [$\bar{z}$ =
a) $80$, b) $81$, c) $83$, d) $84$]. The differential rotation of the macro-particle
is clearly displayed. For the parameters choice refer to the caption of Fig. \ref{fig1}.
\label{fig2}}
\end{figure}

For a monokinetic initial beam with velocity resonant with the wave,
equations  (\ref{eq:mvttheta}), (\ref{eq:mvtp}) and (\ref{eq:mvtA})
predict an exponential instability and a late oscillating
saturation for the amplitude of the radiation field.
Numerical simulations fully confirm this scenario as displayed in
fig.~\ref{fig1}. In the single particle ($q,p$) space, a dense core of particles is trapped by
the wave and behaves like a large ``macro-particle'', that evolves coherently in the
resonance. The distances between these particles do not grow
exponentially fast (as is the case for chaotic motion) but grow at most
linearly with time (for particles trapped in the resonance with different
adiabatic invariants, i.e.essentially different action in the single
particle pendulum-like description). This linear-in-time departure of the
particles appears in the differential rotation in fig.~\ref{fig2}, while
the  remaining particles are almost uniformly distributed between two
oscillating boundaries. Having observed the formation of such structures in the
phase-space allowed to derive a simplified Hamiltonian model to
characterize the asymptotic evolution of the laser \cite{tennyson, andrea}. This reduced
formulation consists in only four degrees of freedom, namely the wave, the macro-particle and the
two boundaries delimiting the portion of space occupied by the
so-called {\it chaotic sea}, i.e.\ the uniform halo surrounding the
inner core.

%%%%%%%%%%%%%%%%%%%%%%%%%%%%%%%%%%%%%%%%

%%%%%%%%%%%%%%%%%%%%%%%%%%%%%%%%%%%%%%%%

%\vspace{1truecm}
%%%%%%%%%%%%%%%%%%%%%%%%%%%%%%%%%
%%%%%%%%%%%%%%%%%%%%%%%%%%%%%%%%%%%%

In \cite{andrea} we hypothesized the macro-particle to be formed
by $N_m$ individual massive units, and
introduced the variables
$(\zeta,\xi)$ to label its position in the phase space.

The $N_c=N-N_m$ particles of the surrounding halo are treated
as a {\it continuum} with  constant phase space distribution,
$f_{sea}(\theta,p,\bar{z}) = f_c$, between two boundaries, namely $p_{+}(\theta,\bar{z})$ and
$p_{-}(\theta,\bar{z})$ such that:

\begin{equation}
\label{eqpmp}
    p_{\pm} = p^{0}_{\pm} + \widetilde{p}_{\pm} \exp(i \theta) + \widetilde{p}_{\pm}^*\exp(-i \theta)~,
\end{equation}
where $p^{0}_{\pm}$ represents their mean velocity. These assumptions allow to map the original
system, after linearizing with respect to $\widetilde{p}_{\pm}$, into \cite{andrea}:

\begin{eqnarray}
    \ddot{\zeta} &=& i \Phi e^{i \zeta} - i \Phi^{\ast} e^{-i
    \zeta} \\
    \frac{1}{2} \dot{V}_{\pm} &=& - \frac{i}{2} {\omega}_{\pm} V_{\pm} + i \Phi \\
 \dot{\Phi} &=& \frac{i}{2} \frac {N_{c}} {N} \frac { V_{+} - V_{-} }  {
    \omega_{+} - \omega_{-} } + i \frac {N_{m}} {N} e^{-i \zeta} + i\delta \Phi
\end{eqnarray}
where \footnote{It is worth stressing that the notation is
  slightly changed with respect to the one adopted in \cite{andrea}, aiming at
  simplifying the forthcoming calculations.}

\begin{eqnarray}
 A & = & - i \Phi \\
 \widetilde {p}_{\pm} & = & V_{\pm} / 2 \\
 \widetilde {p}^{0}_{\pm} & = & \omega_{\pm}
\end{eqnarray}

Normalizing the density in the chaotic sea to unity yields $f_c=1/(2\pi\Delta
\omega)$, where
 $\Delta \omega := \omega_+-\omega_-$ represents the (average) width of the chaotic sea.
The above system can be cast in a Hamiltonian
form by introducing new actions $I_\pm$ and their conjugate angles $\varphi_\pm$:

\begin{eqnarray}
    V_{+} &=& \sqrt {4 \frac{I_{+} \Delta \omega} {N_{c}}} e^{- i \varphi_{+} } \\
    V_{-} &=& \sqrt {4 \frac{I_{-} \Delta \omega} {N_{c}}} e^{i \varphi_{-} }.
\end{eqnarray}

\begin{figure}[t]
\centering
 \includegraphics[width=8 truecm]{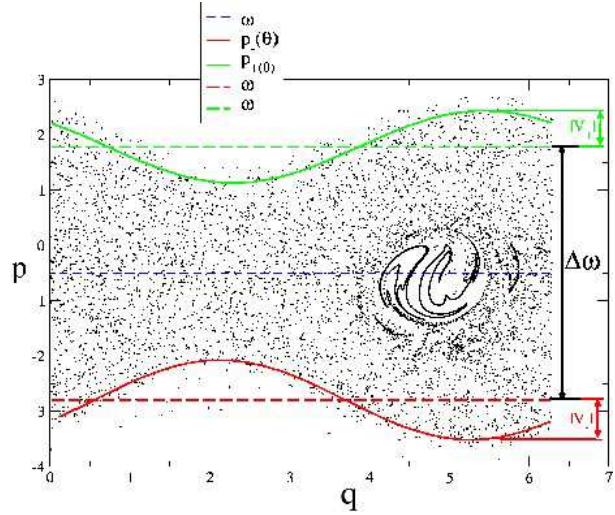}
\caption{\em  ($q$,$p$) phase space portrait in the deep saturated regime for a
monokinetic initial beam ($I(0)
  \simeq  0$, $p_j(0)=0$ and $q$ uniformly distributed in
$[0,2\pi]$).
The two solid lines result from a numerical fit performed according to
the following strategy. First,
the particles located close to the outer boundaries are selected
and then the expression $p_{\pm}(q)=\omega_{\pm} \mp |V_{\pm}| \sin(q+B_{\pm})$ is numerically adjusted to interpolate
their distribution. Here, $|V_{\pm}|$, $B_{\pm}$ and $\omega_{\pm}$ are free parameters. The
numerics are compatible with the simplifying assumption
$B_+=B_- \simeq 0$.
\label{fig3}}
\end{figure}

A pictorial representation of the main quantities
involved in the analysis is displayed in fig. \ref{fig3}. In addition:

\begin{equation}
    \Phi =  i A = - \sqrt {\frac{I} {N}} e^{-i ( \varphi + \frac{\pi}
    {2})}.
\end{equation}
The reduced $4$-degrees-of-freedom Hamiltonian reads, up to a constant
irrelevant to the evolution equations:

\begin{equation}
\label{Ham_fin}
H_4=\frac {\xi^{2}}{2 N_{m}} -\delta I + \omega_{+} I_{+} - \omega_{-} I_{-} - 2
\alpha \left [ \sqrt{I I_{+}} \sin (\varphi - \varphi_{+}) - \sqrt{I
I_{-}} \sin (\varphi + \varphi_{-}) \right ] - 2 \beta \sqrt{I} \sin
(\varphi - \zeta),
\end{equation}
where
\begin{eqnarray}
  \alpha &=& \sqrt {\frac{N_{c}}{ N \Delta \omega}} \\
  \beta &=& \frac{N_{m}}{\sqrt{N}}
\end{eqnarray}
The first four terms represent the kinetic energy of the
macro-particle, the oscillation of the wave and the harmonic contributions associated to the oscillation of
the chaotic sea boundaries. The remaining terms refer to the
interaction energy. Total momentum is $P = \xi + I + N_c \bar\omega$.
The Hamiltonian (\ref{Ham_fin}) allows for a
simplified description of the late nonlinear regime of the
instability, provided the three parameters $\omega_+$, $\omega_-$ and
$N_m$ are given.

To achieve a complete and satisfying theoretical
description we need to provide an argument to self-consistently
estimate these coefficients.  To this end, we shall use
the analytical characterization of the asymptotic behavior of the laser
intensity and beam bunching (a measure of the
  electrons spatial modulation) obtained in \cite{julien} with a
statistical mechanics approach. In the next section these results are
shortly reviewed.

%%%%%%%%%%%%%%%%%%%%%%%%%%%%%%%%%%%%%%%%%%%%%%%%%%%%%%%%%%%%%%%%%%%%%%%%%%%%%%%%%%%%%%%%%%%%%%%%

%%%%%%%%%%%%%%%%%%%%%%%%%%%%%%%%%%%%%%%%%%%%%%%%%%%%%%%%%%%%%%%%%%%%%%%%%%%%%%%%%%%%%%%%%%%%%%%%

%*******************************************************************
\section{Statistical theory of Single-Pass FEL saturated regime}
\label{sec3}
%*******************************************************************

As observed in the previous Section, the process of
wave amplification occurs in two steps: an initial
exponential growth followed by a relaxation towards a
quasi-stationary state characterized by large oscillations. This regime is
governed by the Vlasov equation,  rigorously obtained by
performing the continuum limit ($N \rightarrow \infty$ at fixed volume and
energy per particle) \cite{julien,ElskensBook,Firpo98} on the discrete system
(\ref{eq:mvttheta}-\ref{eq:mvtA}). Formally, the following Vlasov-wave system is found:

\begin{eqnarray}
\frac{\partial f}{\partial \bar{z}} &=& -p\frac{\partial
f}{\partial \theta}
+2(A_x\cos{\theta}-A_y\sin{\theta})\frac{\partial f}{\partial
p}\quad ,
\label{eq:VlasovFELa}\\
\frac{d A_x}{d \bar{z}}   &=& -\delta
A_y+ \int f \cos{\theta}  \, {\mathrm d}\theta \,
{\mathrm d}p\quad ,
\label{eq:VlasovFELb}\\
\frac{d A_y} {d \bar{z}}  &=&  \delta
A_x- \int f \sin{\theta} \, {\mathrm d}\theta \,
{\mathrm d}p\quad . \label{eq:VlasovFELc}
\end{eqnarray}
The latter conserves the pseudo-energy per particle

\begin{equation}\label{vlasovenergy}
\epsilon= \int{\frac{p^2}{2}f(\theta,p)\,{\mathrm d}\theta \,{\mathrm d}p}
 - \delta (A_x^2 + A_y^2) + 2\int{ (A_x\sin\theta +A_y\cos\theta)f(\theta,p)}  \,{\mathrm d}\theta \,
{\mathrm d}p
\end{equation}
and the momentum per particle

\begin{equation}\label{vlasovmomentum}
\sigma= \int{p f(\theta,p)\,{\mathrm d}\theta \,{\mathrm d}p}
 + (A_x^2 + A_y^2)  \  .
\end{equation}

A subsequent slow relaxation
towards the Boltzmann equilibrium is observed. This is a typical finite-$N$ effect
and  occurs on time-scales much longer than the transit trough the undulator \cite{ElskensBook,Firpo01,Bonifacio}. For our calculations we are
interested in the first saturated state.
To estimate analytically the average
intensity and bunching parameter in this regime
we exploit the statistical treatment of the Vlasov
equation, presented in \cite{julien}.
In the following, we
provide a short outline of the strategy.
Since the Gibbs ensembles are equivalent for this model,
note that the same expressions are recovered through a canonical calculation \cite{ElskensBook, julien,
Firpo00}.

The basic idea is to coarse-grain the microscopic one-particle distribution
function $f(\theta,p,\bar{z})$. An entropy is then associated to the
coarse-grained distribution $\bar{f}$, which essentially counts a number of
microscopic configurations. Neglecting the contribution of the field,
since it represents only one degree of freedom within the ($N+1$) of
the Hamiltonian (\ref{eq:Hamiltonien}), one assumes

\begin{equation}\label{vlasoventropy}
  s(\bar{f})
  =
-\int{\left[\frac{\bar{f}}{f_0}\ln \frac{\bar{f}}{f_0}
    +\left(1-\frac{\bar{f}}{f_0}\right) \ln\left(1-\frac{\bar{f}}{f_0}\right)\right]
    f_0 {\mathrm d}\theta \,{\mathrm d}p}
  \simeq
-\int \left[\frac{\bar{f}}{f_0}\ln \frac{\bar{f}}{f_0}\right] f_0 \dd \theta \dd p \  ,
\end{equation}
where  the constant $f_0$ is related to the initial distribution
  \footnote{More generally, we require $f_0$ to be a reference number with
    appropriate dimension, chosen so large that the entropy reduces to
the last expression in the case of interest to us.}.

The equilibrium is computed by maximizing this
entropy, while imposing the dynamical constraints. This corresponds to
solving the constrained variational problem
\begin{equation}\label{valsoventropy2}
S(\epsilon,\sigma)=\max_{\bar{f},A_x,A_y}\left(s(\bar{f})\Big|H(\bar{f},A_x,A_y)=N\epsilon;
P(\bar{f},A_x,A_y)=N\sigma;\int{ f(\theta,p)\,{\mathrm d}\theta \,{\mathrm
d}p}=1\right),
\end{equation}
which leads to the equilibrium values

\begin{eqnarray}
\label{equilibriumvalues}
\bar{f} &=& f_0 \frac{e^{-\beta(p^2/2+2A \sin
    \theta)-\lambda p -\mu}}{1+e^{-\beta(p^2/2+2A\sin \theta) - \lambda p -\mu}}\\
A&=&\sqrt{A_x^2+A_y^2}=\frac{\beta}{\beta \delta-\lambda}\int{\sin(\theta)\bar{f}(\theta,p)\,{\mathrm
d}\theta \,{\mathrm d}p},
\end{eqnarray}
where $\beta$, $\lambda$ and $\mu$ are the Lagrange multipliers for the energy,
momentum and normalization constraints and, in addition, we have assumed the
non-restrictive condition $\sum \cos(\theta_i)=0$ \cite{julien}. Using then the three
equations for the constraints, the statistical equilibrium
calculation is reduced to finding the values of the multipliers as functions
of energy $\epsilon$ and momentum $\sigma$. These equations lead directly to the estimates
of the equilibrium values for the intensity $I$ and bunching parameter
$b=|\sum \exp(i \theta_i)|/N$.

\begin{figure}[ht]
 \resizebox{0.64\textwidth}{!}{\includegraphics{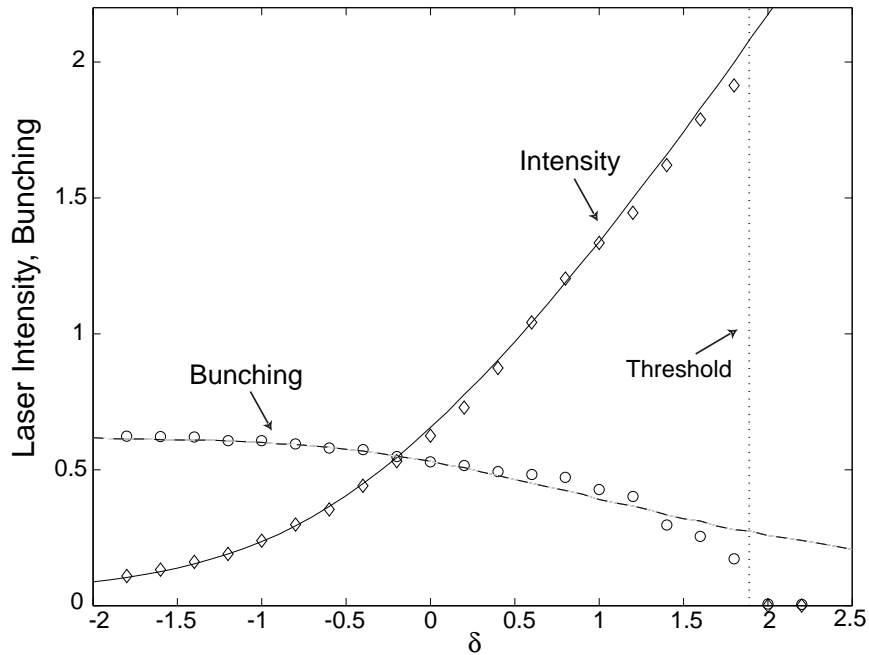}}
\vskip .5truecm \caption{\em
 Comparison between theory (solid
and long-dashed lines) and simulations (symbols) for a monoenergetic beam with $\epsilon = 0$, $\sigma = 0$,
when varying the detuning $\delta$. The dotted vertical line,
$\delta=\delta_c\simeq1.9$,  represents the transition from the high-gain to
low gain regime \cite{Bonifacio}.
\label{figSTAT}}
\end{figure}

In the following, we focus on the case of an initially monokinetic beam injected
 at the wave velocity, while the initial wave intensity is negligible, so that $\epsilon = 0$
 and $\sigma = 0$. Moreover we let $f_0 \rightarrow \infty$, which amounts to
 $\mu \rightarrow \infty$ in eq. (\ref{equilibriumvalues}). Results are displayed
  in fig.~\ref{figSTAT} showing remarkably good
agreement between theory and simulations,
below the critical threshold
$\delta_c \simeq 1.9$ that marks the transition between high and low
 gain regimes. This transition is purely dynamical and cannot be
 reproduced by the statistical calculation.

Analytically, it turns out that
\begin{equation}
  b = |A|^3 - |A| \delta
    = \frac {{\mathrm{I}_1}(2/(3 |A|^3 - 2 \delta |A|) ) } {{\mathrm{I}_0}(2/(3 |A|^3 - 2 \delta |A|) ) }
  \label{eq:b}
\end{equation}
where ${\mathrm{I}}_n$ is the modified Bessel function of order $n$. In particular for $\delta=0$,
one finds $|A|^2 = I/N \simeq 0.65$ and $b \simeq 0.54$.

%%%%%%%%%%%%%%%%%%%%%%%%%%%%%%%%%%%%%%%%

%%%%%%%%%%%%%%%%%%%%%%%%%%%%%%%%%%%%%%%%

%*******************************************************************
\section{Towards the analytical characterization of $N_m$}
\label{sec4}
%*******************************************************************

As previously discussed, one can predict the value of the bunching
parameter $b$, using the above statistical mechanics description. Clearly,
the bunching parameter $b$ depends on the spatial distribution
of the particles. From its definition it immediately follows:

\begin{equation}
\label{bunch_espr}
b=\left[ \left( \frac{1}{N} \sum_{i=1}^N \cos q_i \right)^2 +
\left( \frac{1}{N} \sum_{i=1}^N \sin q_i \right)^2 \right]^{1/2}.
\end{equation}

To proceed we can isolate the contribution relative to the
macroparticle from that associated to the chaotic sea.
We thus obtain:

\begin{equation}
\label{bunch_espr2}
b=\left[\left(
\frac{1}{N} \sum_{i \in macro} \cos q_i
+ \frac{1}{N} \sum_{i \in sea} \cos q_i
\right)^2 +
\left(
\frac{1}{N} \sum_{i \in macro} \sin q_i
+ \frac{1}{N} \sum_{i \in sea} \sin q_i
\right)^2 \right]^{1/2}.
\end{equation}

Focus on the first two terms of expression (\ref{bunch_espr2}).
We assume the macroparticle to be ideally localized at
 the elliptic fixed point  solving (\ref{pfix})-(\ref{qfix}), i.e.\ set
$q_j=\bar{q}=3\pi / 2$, for each individual massive unit belonging to the
inner agglomeration. Hence,
$\sum_{i \in macro} \cos q_i=N_m \cos \bar{q} = 0$.
As concerns the second contribution, recalling the expressions for the boundaries
$p_-(q)$ and  $p_+(q)$ (see caption of fig. \ref{fig3}), one can formally write:

\begin{eqnarray}
\label{bunch_cos_sea}
  \frac{1}{N} \sum_{i \in sea} \cos q_i
  &\simeq&
  \frac{N_c}{N}\frac{1}{2\pi(\omega_+ - \omega_-)}
    \int_{0}^{2\pi} \cos q \int_{p_-(q)}^{p_+(q)}\dd p \dd q
  \nonumber \\
  &=&
  \frac{N_c}{N}\frac{1}{2\pi(\omega_+ - \omega_-)}
    \int_{0}^{2\pi} (\cos q)[\omega_+ -\omega_-]\dd q
  \nonumber \\
  &&
  -\frac{N_c}{N}\frac{1}{2\pi(\omega_+ - \omega_-)}
    \int_{0}^{2\pi} [(|V_+|+|V_-|)\sin q]\cos q \dd q=0.
\end{eqnarray}

The other contributions in eq. (\ref{bunch_espr2}) can be estimated as follows:

\begin{equation}
\label{bunch_sin_macro}
  \frac{1}{N} \sum_{i \in macro} \sin q_i
  \simeq
  -\frac{N_m}{N}
\end{equation}
while

\begin{eqnarray}
  \label{bunch_sin_sea}
  \frac{1}{N} \sum_{i \in sea} \sin q_i
  &\simeq&
  \frac{N_c}{N}\frac{1}{2\pi(\omega_+ - \omega_-)}
    \int_{0}^{2\pi} (\sin q) \int_{p_-(q)}^{p_+(q)}\dd p \dd q
  \nonumber \\
  &=&
  \frac{N_c}{N}\frac{1}{2\pi(\omega_+ - \omega_-)}
      \underbrace{\int_{0}^{2\pi} \sin q (\omega_+ - \omega_-)\dd q}_{=0}
  \nonumber \\
  &&
  -\frac{N_c}{N}\frac{1}{2\pi(\omega_+ - \omega_-)}\int_{0}^{2\pi}
(|V_+| + |V_-|)\sin^2 q \dd q
  = -\frac{N_c}{N}\frac{\widetilde{V}}{2\Delta \omega},
\end{eqnarray}
where $\widetilde{V}=|V_+|+|V_-|$.
Inserting (\ref{bunch_cos_sea}), (\ref{bunch_sin_macro}), (\ref{bunch_sin_sea}) into (\ref{bunch_espr2}) yields:

\begin{equation}
  \label{bunch_final}
  b=\frac{N_m}{N} + \frac{N_c}{N}\frac{\widetilde{V}}{2\Delta\omega}
\end{equation}
and  finally

\begin{equation}
  \label{N_m}
  \frac{N_m}{N}
  = \frac{2 b \Delta \omega - \widetilde{V}}{2 \Delta \omega-\widetilde{V}} =
  1 -\frac{2(1-b)\Delta \omega}{2\Delta \omega - \widetilde{V}}
\end{equation}
using
the relation $N_m=N-N_c$. It is worth
emphasizing that, neglecting as a first approximation the amplitudes
of the sinusoidal boundaries, i.e. setting $V_-=V_+=0$, the previous
equation reduces to

\begin{equation}
\label{N_m1}
  \frac{N_m}{N} = b~,
\end{equation}
confirming the relevance of the macroparticle picture to the bunching parameter, a physical quantity of paramount importance
for the FEL dynamics. Formula (\ref{bunch_final}) yields bunching parameter values
larger than (\ref{N_m1}), i.e. implies that the sea also contributes to increasing the bunching parameter. The difference increases
as the sea gets more populated and the width $\omega_+ - \omega_-$ is
reduced (note that our approximations require $|V_+| + |V_-|< \omega_+ - \omega_-$,
see fig.~\ref{fig3}).

%*******************************************************************
\section{Estimating the average momentum of the boundaries $\omega_{\pm}$}
\label{sec5}
%*******************************************************************

In this section we estimate the unknown quantities
$\omega_{\pm}$ by characterizing their functional
dependence on $N_m$. For this purpose we introduce
(see schematic layout of fig.~\ref{fig3}):

\begin{equation}
\omega_{\pm} = \bar\omega \pm \frac{\Delta\omega}{2}~.
\end{equation}

\begin{figure}[t]
\vspace{2truecm}
\centering
 \includegraphics[width=8 truecm]{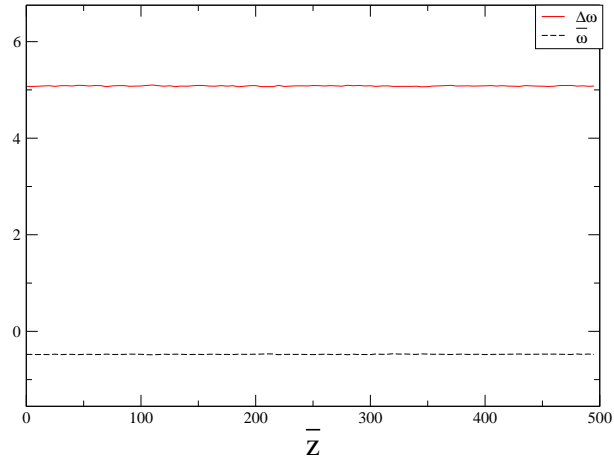}
\caption{\em Solid line: $\Delta \omega$ vs $\bar{z}$. Dashed line: $\bar\omega$
vs $\bar z$.
Parameters are set as discussed in the caption of fig. \ref{fig1}.
\label{fig5}}
\end{figure}

The problem of
estimating $\omega_{\pm}$ is obviously equivalent to providing a
self-consistent calculation for $\bar\omega$ and $\Delta \omega$.
The latter are both monitored as function of
time in fig.~\ref{fig5} and shown to be practically constant.

In the following, we shall focus on the case of a system which evolved from
an initially monokinetic beam and an initially infinitesimal wave. It is
then convenient to choose the Galilean reference frame moving at the beam
initial velocity. This translates into the conditions $\epsilon = 0$ and
$\sigma = 0$. The detuning $\delta$ is arbitrary so far. It is of prime interest to
consider the special case where the beam is injected at the resonant
velocity, so that $\delta = 0$. We shall make this additional
assumption in section \ref{sec7}, but the estimates in this section and in the next one
do not require it unless explicitly
stated.

%%%%%%%%%%%%%%%%%%%%%%%%%%%%%%%%%%%%%%%%%%%%%%%%%%%%%%%%%%%%%%%%%%%%%%%%%%%%%%%%%%%%%%%%%%%%%%%%

%%%%%%%%%%%%%%%%%%%%%%%%%%%%%%%%%%%%%%%%%%%%%%%%%%%%%%%%%%%%%%%%%%%%%%%%%%%%%%%%%%%%%%%%%%%%%%%%

Consider the {\it conservation of momentum} for the original $N$-body system
(\ref{eq:Hamiltonien}) and focus on the asymptotic dynamics, which allows one to
isolate the contributions respectively associated to the macroparticle and the chaotic sea.
 Averaging over the number of particles yields:

\begin{equation}
\label{cons_mom_1a}
  \frac{N_c}{N} p_{sea} + \frac{N_m}{N} p_{macro} + J = \sigma_0
\end{equation}
where $p_{sea}$ stands for the average momentum of the chaotic
sea and the subscript $'0'$ labels the initial condition.
To simplify the calculations, we introduced the rescaled intensity
$J=I/N$. As already observed in \cite{andrea,tennyson}, the macroparticle
rotates in phase space. This rotation is directly coupled to the
oscillations displayed by the laser intensity. Averaging over a bounce period $z_{rot}$, one
formally gets:

\begin{equation}
\label{cons_mom_1b}
  \frac{N_c}{N} \langle p_{sea}\rangle
  +  \frac{N_m}{N} \langle p_{macro} \rangle
  + \langle J \rangle
  = 0
\end{equation}
where $\langle \cdot \rangle$ stands for the time average.
Focus now on $\langle p_{sea}\rangle$. Since particles are uniformly
filling the chaotic sea, one can use the approximation outlined before eq.
(\ref{eqpmp}) (see also caption of fig.\ref{fig3}):

\begin{equation}
\label{mom_sea}
  \langle p_{sea}\rangle
  =
  \frac{1}{z_{rot}}\int_{{\bar
 z}_0}^{{\bar{z}}_0+z_{rot}}
    \left(\int\!\!\int{p f_{sea} (\theta,p,\bar{z}) \dd \theta \dd p}\right) \dd \bar{z}
  =  \frac{1}{2\pi(\omega_+ - \omega_-)} \int \int_{p_-(q)}^{p_+(q)}{p}
 \dd p \dd \theta
  \simeq (\omega_+ + \omega_-)/2
  = \bar \omega~.
\end{equation}

As already outlined in the preceding discussion, we assume that the macroparticle oscillates
 around the fixed point and therefore each individual element constituting the
macroparticle verifies the condition $\langle q_j \rangle =\bar{q}$.
In addition, from equation (\ref{pfix}):

\begin{equation}
\label{p}
  \langle p_{macro} \rangle
  := \bar{p}
  =  \left \langle \frac{1}{N\sqrt{J}} \sum_{i \in   macro} \sin q_i \right \rangle  + \left \langle \frac{1}{N\sqrt{J}} \sum_{i \in
  sea} \sin q_i \right \rangle - \delta.
\end{equation}

To proceed in the analysis, we approximate the right hand side in
equation (\ref{p}) as:

\begin{equation}
\label{approx}
 \left \langle \frac{1}{N\sqrt{J}} \sum_{i \in
  macro} \sin q_i \right \rangle \simeq \frac{1}{N\sqrt{\langle J \rangle}} \sum_{i \in
  macro} \sin \langle q_i \rangle
  = \frac{1}{N\sqrt{\langle J \rangle}} \sum_{i \in  macro} \sin \bar{q}
  = -\frac{N_m}{N} \frac{1}{\sqrt{\langle J \rangle}}
\end{equation}
consistently with the argument after (\ref{bunch_espr2}).
The above relation is derived by performing a linearization (see
Appendix), validated numerically  and
supported a posteriori by the correctness of the results.
The contribution of the chaotic sea reads:
\begin{eqnarray}
\label{sea_contr}
 \left \langle \frac{1}{N\sqrt{J}} \sum_{i \in
  sea} \sin q_i \right \rangle \simeq
\frac{1}{N\sqrt{\langle J \rangle}} \sum_{j\in sea} \sin q_j \simeq \frac{N_c}{N\sqrt{\langle J \rangle}}\frac{1}{2\pi(\Delta\omega)}
\int_0^{2\pi} \sin q \int_{p_-(q)}^{p_+(q)} \dd p \dd q \nonumber \\
= \frac{N_c}{N\sqrt{\langle J
\rangle}}\frac{1}{2\pi\Delta\omega}\left[\underbrace{
\int_0^{2\pi}(\omega_+ -\omega_-)\sin q \dd q}_{=0} - \int_0^{2\pi}(|V_+|+|V_-|)\sin^2 q \dd q  \right]
 =  -\frac{N_c}{N\sqrt{\langle J \rangle}}\frac{\widetilde{V}}{2\Delta\omega}.
\end{eqnarray}

Merging equations (\ref{p}), (\ref{approx}) and (\ref{sea_contr}) and recalling (\ref {bunch_final}) and (\ref{eq:b}), one obtains

\begin{equation}
\label{p_1}
  \langle p_{macro} \rangle
  \simeq
  - \frac{N_m}{N} \frac{1}{\sqrt{\langle J \rangle}}
  -  \frac{N_c}{N}\frac{\widetilde{V}}{2\Delta\omega} \frac{1}{\sqrt{\langle J
\rangle}} - \delta =
  - \frac {b} {\sqrt{\Jav}} - \delta = -\Jav.
\end{equation}
Thus the macroparticle moves on the average at the same velocity as the center of the chaotic sea.
Inserting equations (\ref{mom_sea}) and (\ref{p_1}) into
(\ref{cons_mom_1b}) and solving for $\bar{\omega}$, we find,

\begin{equation}
\label{mean_omeg}
  \bar{\omega}
  =
  \frac{N}{N_c}\left[-\langle J \rangle + \frac{N_m}{N} \left(\frac{b}{ \sqrt{\langle J \rangle}} + \delta \right) \right]
  = - \Jav   \ .
\end{equation}

To get an expression for $\Delta \omega$, we consider the {\it energy
conservation} for the original $N$-body  model (\ref{eq:Hamiltonien}). By averaging over one complete rotation of the macroparticle, we write:

\begin{equation}
\label{cona_en_1a}
\frac{1}{N}\frac{\langle \sum_{i=1}^{N} p_i^2 \rangle}{2} + 2 \left \langle \sqrt{J} \frac{1}{N}
\sum_{i=1}^N  \sin q_i \right \rangle  = \delta \Jav~.
\end{equation}

We then bring into evidence the contributions associated to the
massive agglomerate and to the particles of the surrounding halo,
for both the kinetic and the interaction terms:

\begin{equation}
\label{cona_en_1b}
\frac{N_m}{N}\frac{\langle p_{macro}^2 \rangle}{2} + \frac{N_c}{N}\frac{\langle p_{sea}^2 \rangle}{2}
+ \left \langle  2 \sqrt{J} \frac{1}{N} \sum_{i \in
  macro} \sin q_i \right \rangle  + \left \langle  2 \sqrt{J} \frac{1}{N} \sum_{i \in
  sea} \sin q_i \right \rangle = \delta \Jav~.
\end{equation}

Hereafter we make use of
$\langle p_{macro}^2 \rangle \simeq
\langle p_{macro} \rangle^2$, which in turn amounts to
assume small oscillations around the mean $\langle p_{macro} \rangle$,
consistently with (\ref{p}) which neglects such oscillations.
The kinetic energy associated to the uniform sea
can be estimated as follows:

\begin{equation}
\label{en_sea_1}
 \frac{\langle p_{sea}^2\rangle}{2} = \frac{1}{2\pi(\omega_+ - \omega_-)}\int{\dd \theta\int_{\omega_-}^{\omega_+}{\frac{p^2}{2}}
 \dd p} = \frac{1}{6}\frac{\omega_+^3-\omega_-^3}{\omega_+ - \omega_-} = \frac{1}{6}\left(3\bar{\omega}^2 + \frac{\Delta\omega^2}{4}\right).
\end{equation}

In this estimate we assumed the particles to be distributed uniformly in a rectangular
box, disregarding the sinusoidal shape of the boundaries. The modulation of the outer frontiers results
in higher order corrections which can be neglected.

The interaction term follows directly from (\ref{p}) and (\ref{p_1}). Inserting (\ref{en_sea_1}) in (\ref{cona_en_1b}), one finds:

\begin{equation}
\label{deltaomega}
\Delta \omega^2
  =  \frac{N}{N_c} \left[ 36 \Jav^2 - 24 \delta \Jav  \right]  .
\end{equation}

%*******************************************************************
\section{Closed expressions for the  amplitudes $V_{\pm}$ of the outer boundaries}
\label{sec6}
%*******************************************************************

The preceding calculations lead to three equations that allow to
estimate $N_m$, $\bar{\omega}$ and $\Delta \omega$ (and thus $\omega_+$ and
$\omega_-$), provided an expression for $V_{\pm}$ is given.

Note that $V_{\pm}$ enter as variables in the
Hamiltonian and thus evolve self-consistently. However, one can assess their average value using
 an adiabatic argument \cite{ElskensBook}.
The boundaries of the sea are rather sharply drawn by the motion of particles, which move following the time-dependent Hamiltonian
$H_{1.5} = p^2/2 + 2 \sqrt{J} \sin (\theta - \varphi)$, with 1.5 degrees of freedom. In a first approximation, the time dependence of $J$ and $\varphi$ results in a detuning of the wave, as
\begin{equation}
  \dot \varphi
  = \partial_I H
  = - \delta + (NI)^{-1/2} \sum_j \sin (\theta_j - \varphi)
  = - \delta - b / \sqrt{J}
  = - J
\end{equation}
where we used (\ref{eq:b}) and the estimates of sec.~\ref{sec4}. The resulting velocity agrees with the canonical estimate of reference \cite{ElskensBook} in the low-temperature regime, as it must since ensembles are equivalent for this model.

Let us neglect the pulsations of $J$ and further variations of $\dot \varphi$.
Then the Hamiltonian $H_{1.5}$ is brought to the integrable form
of the classical pendulum, by a Galileo transformation
to the frame moving at velocity $v = - \Jav$.
In this frame, the motion of particles preserves their action,
which is directly related to their effective energy

\begin{equation}
\label{ad_arg1}
  E_{\pm}
  =
  \frac{(p'_{\pm})^2}{2}
  + 2 \sqrt{\langle J \rangle} \sin q'
\end{equation}
where $p'_\pm = p_\pm + \Jav$, $q' = \theta + \Jav t - \varphi'$ ;
here $\varphi' = \varphi + \Jav t$ is the wave phase in the new frame. Solving for $p$, we get:

\begin{eqnarray}
\label{ad_arg2}
  p_{\pm}  + \Jav
  & = &
  \pm \sqrt{2E_{\pm}}\sqrt{1 - \frac{2\sqrt{\langle J \rangle}}{E_{\pm}}\sin q'}
  \simeq
    \pm \sqrt{2E_{\pm}}\left(1 - \frac{\sqrt{\langle J \rangle}}{E_{\pm}}\sin q'\right) \nonumber \\
& \simeq & (\omega_{\pm} + \Jav) \mp |V_{\pm}|\sin q' = \pm \left( \frac{\Delta
\omega}{2} - |V_{\pm}|\sin q'  \right)
\end{eqnarray}
where we used the assumption that $\Jav^{1/2} / E_\pm$ is small, i.e.\ that the particle motion near the sea boundary (which has period $\sim 2 \pi / \sqrt{2 E_\pm}$) is fast with respect to the characteristic frequency $\sqrt{2} \langle J \rangle^{1/4}$ of particle oscillations inside the wave potential well.
From (\ref{ad_arg2}) we finally obtain

\begin{equation}
\label{expr_V}
  |V_{\pm}|
  =
   \frac{4\sqrt{\langle J \rangle}}{\Delta \omega}~.
\end{equation}

Note that the boundaries found here are also approximately level lines of the
distribution $\bar f$ given by eq. (\ref{equilibriumvalues}), as they should be.
%*******************************************************************
\section{Validating the theory through direct simulation}
\label{sec7}
%*******************************************************************

In conclusion, we obtained a system of
equations that can be solved
to compute the values of $V_{+}$ and $V_{-}$ and provide self-consistent
estimates of the three quantities $N_m$,
$\omega_+$ and $\omega_-$, through a direct calculation of
$\bar{\omega}$ and $\Delta \omega$,
for an initially monokinetic beam and infinitesimal wave ($H/N=0$,
$P/N=0$). The final equations read in the case $\delta = 0$:

\begin{eqnarray}
  \label{finalsystem}
  \frac{N_m}{N}
    &=& \frac{b  - \widetilde{V}/(2\Delta \omega)}
             {1  - \widetilde{V}/(2\Delta \omega )}
  \nonumber \\
  \bar{\omega}
  &=&
  - \Jav
  \\
  \Delta \omega
  &=&
  6 \sqrt{\frac{N}{N_c}} \Jav
  \nonumber \\
  |V_\pm|
  &=&
    \frac{4\sqrt{\langle J \rangle}}{\Delta \omega}
  \nonumber \\
  \widetilde{V}
  &=&
  |V_+| + |V_-|
  \nonumber
\end{eqnarray}
where $\Jav = |A|^2$ and $b$ are calculated from (\ref{eq:b}).
The system is explicit after a few algebraic calculations, which show that
\begin{equation}
  \label{Ncfinal}
  {\frac {N_c} N}
  =
  {9 \over 2} b - \sqrt{{117 \over 4} b^2 - 9 b}  \ .
\end{equation}

It is worth recalling that the derivation of this system assumes that
the dynamical variables like $V_\pm$ and $J$ do not vary too much,
so that these final estimates yield coefficients $\omega_\pm$
which do not depend on time.

In table \ref{tab3} the
solutions of system (\ref{finalsystem}) are compared with the values
measured in direct numerical experiments.

\begin{figure}[t]
\vspace{1truecm}
\centering
 \includegraphics[width=8 truecm]{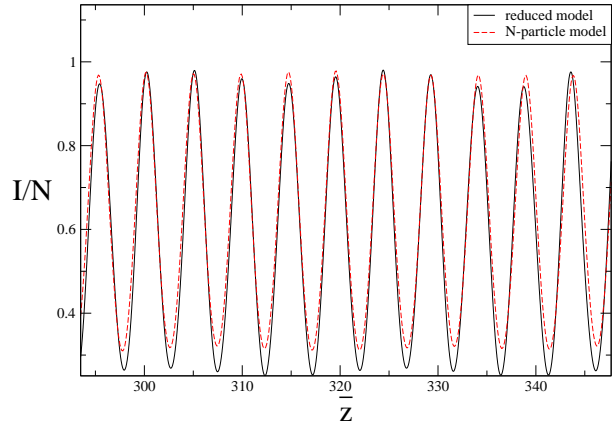}
\caption{\em Rescaled laser intensity I/N vs time. The solid line refers to the
  reduced dynamics. The free parameters are fixed according to the
  self-consistent derivation here discussed. The dashed
  line refers to a direct integration of the original $N$-body system (\ref{eq:Hamiltonien}).
\label{fig6}}
\end{figure}

\begin{figure}[t]
\vspace{1truecm}
\centering
 \includegraphics[width=8 truecm]{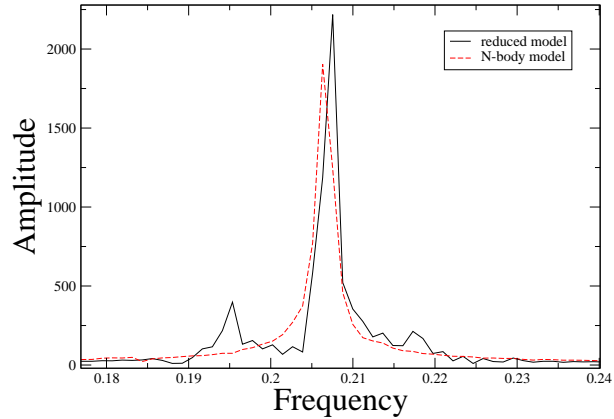}
\caption{\em Fourier transform of the signals reported in fig.~\ref{fig6}. The thick solid line refers to the
reduced model, while the dashed one refers to the original $N$-body system.
\label{fig7}}
\end{figure}

The theoretical values are then used to simulate the
reduced dynamics (\ref{Ham_fin}). Comparisons with numerical results
based on the original $N$-body (\ref{eq:Hamiltonien}) model are reported in
fig.~\ref{fig6} and display a remarkably good agreement. The normalized standard
 deviations of the intensity in the two cases differ by less than five percent.
As an additional check, we report in fig.~\ref{fig7} the Fourier transform of the signals
displayed in fig.~\ref{fig6}. A discrepancy smaller than $1$~\%  is
observed for the peak values. These results provide an a posteriori validation of
the reduced model which can be effectively employed to characterize the
saturated evolution of the FEL field.

\begin{table}[h!]
\begin{center}
\vspace{5mm}
\begin{tabular}{ |c| |c| |c|}
\hline
&Theory&Numerics\\
\hline
\hline
$b$ & $0.530$ & $0.56$\\
\hline
$\langle J \rangle$ & $0.655$ & $0.62$\\
\hline
$N_m/N$ & $0.472$ & $0.5$\\
\hline
$\bar{\omega}$ & $-0.655$ & $-0.50$\\
\hline
$\Delta\omega$ & $5.408$ & $5.10$ \\
\hline
$|V_+|$ & $0.599$ & $0.85$\\
\hline
$|V_-|$ & $0.599$ & $0.8$ \\
\hline
\end{tabular}
\end{center}
\caption{\label{tab3} \em  Theoretical estimates from system
  (\ref{finalsystem}) (first column) vs. numerical measurements (second
  column) for $\delta = 0$. To estimate $N_m$ we
  select the $i^{th}$ particle and calculate the spatial distances from the
  adjacent neighbors. If such quantities remain below a
  cut-off threshold during the subsequent evolution, the particles are
  said to belong to the inner cluster.}
\end{table}

%%%%%%%%%%%%%%%%%%%%%%%%%%%%%%%%%%%%%%%%%%%%%%%%%%%%%%%%%%%%%%%%%%%%%%%%%%%%%%%%%%%%%%%%%%%%%%%%

%%%%%%%%%%%%%%%%%%%%%%%%%%%%%%%%%%%%%%%%%%%%%%%%%%%%%%%%%%%%%%%%%%%%%%%%%%%%%%%%%%%%%%%%%%%%%%%%

%%%%%%%%%%%%%%%%%%%%%%%%%%%%%%%%%%%%%%%%%%%%%%%%%%%%%%%%%%%%%%%%%%%%%%%%%%%%%%%%%%%%%%%%%%%%%%%%

%%%%%%%%%%%%%%%%%%%%%%%%%%%%%%%%%%%%%%%%%%%%%%%%%%%%%%%%%%%%%%%%%%%%%%%%%%%%%%%%%%%%%%%%%%%%%%%%

\vspace{2truecm}

%*******************************************************************
\section{Conclusion}
\label{sec8}
%*******************************************************************

In this paper we consider a reduced Hamiltonian formulation to
reproduce the saturated regime of a single-pass Free Electron Laser
around perfect tuning. The model consists of only four degrees of
freedom. The particles trapped in the resonance give
birth to a coherent structure localized in space which is formally
treated as an individual macro-particle. The remaining
particles constitute the so-called {\it chaotic sea} and are
uniformly distributed in phase space between two sharp boundaries that
evolve according to the Vlasov equation.

This minimalistic approach
was first derived by Tennyson et al. working in the context of plasma
physics and recently adapted to the FEL in \cite{andrea}.
The original formulation assumes three external parameters, namely the
mass of the dense core, $N_m$, and the average momenta
$\omega_\pm$ associated to the outer boundaries, which have been
so far phenomenologically adjusted to reproduce the results of
$N$-body simulations. The lack of a self-consistent characterization
has significantly weakened the potential impact of the
reduced Hamiltonian approach so far.

This long standing problem
\cite{tennyson,ElskensBook} is here addressed and solved. To this end
we make use of the analytical estimates of the average
intensity and bunching parameter at the saturation based on a statistical mechanics treatment
of the Vlasov equation and of adiabatic theory \cite{ElskensBook, julien}.
Our theoretical predictions for the parameters
are compared to direct numerical measurements, showing an excellent
agreement. Simulations of the reduced dynamics, complemented with
the strategy here discussed, reproduce remarkably well the oscillatory
regime displayed by the original $N$-body model.

In conclusion,
we have provided a solid self-consistent formulation for the reduced
Hamiltonian, which we hereby propose as an alternative theoretical
tool to investigate the saturated regime of a FEL.

%*******************************************************************
\section{Acknowledgments}
%*******************************************************************

We thank R.~Bachelard, C.~Chandre and G.~De~Ninno for fruitful
discussions. D.F. thanks Edison Giocattoli Firenze for financial
support. Y.E. benefited from a delegation position at CNRS.
This work is part of the PRIN2003 project {\it Order and
chaos in nonlinear extended systems} funded by MIUR-Italy.

%*******************************************************************
\section*{Appendix}\label{appA}
%*******************************************************************

Consider the following expression:
\begin{equation}
\label{W}
W=  \frac{1}{N\sqrt{J}} \sum_{i \in macro} \sin q_i
\end{equation}
and suppose we want to evaluate $\langle W \rangle$. The asymptotic intensity
oscillates around a given plateau $\langle J \rangle$, hence we write the
dominant Fourier term (see fig. \ref{fig7}):

\begin{equation}
\label{I}
J=\langle J \rangle \left( 1+ \epsilon_J \sin(\omega \bar{z}) \right)
\end{equation}
with $\epsilon_J=\Delta J / \langle J \rangle$, noting $\Delta J$ the
amplitude of the oscillations. Moreover the macroparticle displays
regular oscillations, centered around $\bar{q} = 3 \pi /2$. This in
turn allows us to write:

\begin{equation}
\label{q}
q_i=\bar{q} + \epsilon_q \sin(\omega \bar{z})
\end{equation}
for the $i^{th}$ particle composing the dense core.  Here $\epsilon_q$
measures the average distance from  $\bar{q}$. Assuming both
$\epsilon_J$ and $\epsilon_q$ to be small, one can approximate
(\ref{W}) as follows:

\begin{equation}
\label{W1a}
W =
  \frac {N_m} {N \sqrt{\Jav}}
    \left(-1 + \frac {\epsilon_q^2} {2} \sin^2 (\omega \bar z) + \ldots \right)
  \left(1 - \frac {\epsilon_J} {2} \sin (\omega \bar z) + \ldots \right)
  \simeq - \frac{N_m}{N\sqrt{\langle J \rangle}} +
\frac{1}{2} \frac{N_m}{N \langle J \rangle^{1/2}} \epsilon_J \sin(\omega \bar{z})
\end{equation}
where use has been made of the fact the $\cos{\bar{q}}=0$.
Averaging (\ref{W1a}) over one period $z_{rot}=2 \pi /\omega$ one gets:

\begin{equation}
\label{W1b}
\langle W \rangle = - \frac{N_m}{N\sqrt{\langle J \rangle}}
\end{equation}
which is identical to equation (\ref{approx}).

\end{document}